\documentclass[12pt]{article}

\usepackage{epsfig}
\usepackage{cite}
\usepackage{amsmath, amssymb, amsfonts}
\usepackage{color}
\usepackage{latexsym}  
\usepackage{graphicx}
\usepackage{cancel}
\usepackage[colorlinks,bookmarks]{hyperref}
\hypersetup{pdfpagemode=UseNone, pdfstartview=FitH, linkcolor=blue, citecolor=red, urlcolor=blue}

\def\be{\begin{equation}}
\def\ee{\end{equation}}
\def\ba{\begin{eqnarray}}
\def\ea{\end{eqnarray}}

\def\bdm{\begin{displaymath}}
\def\edm{\end{displaymath}}
\def\la{~\mbox{\raisebox{-.6ex}{$\stackrel{<}{\sim}$}}~}

\def\bq{\begin{quote}}
\def\eq{\end{quote}}

 at 10truept

\newcommand{\Mpl}{M_{\mathrm{Pl}}}

\newcommand{\bea}{\begin{eqnarray}}
\newcommand{\eea}{\end{eqnarray}}

\newcommand{\bi}{\begin{itemize}}
\newcommand{\ei}{\end{itemize}}

\newcommand{\beq}{\begin{equation}}
\newcommand{\eeq}{\end{equation}}
\newcommand{\beqa}{\begin{eqnarray}}
\newcommand{\eeqa}{\end{eqnarray}}
\newcommand{\mpl}{\Mpl}

\def\12{{1 \over 2}}

\def\ltap{\ \raise.3ex\hbox{$<$\kern-.75em\lower1ex\hbox{$\sim$}}\ }
\def\gtap{\ \raise.3ex\hbox{$>$\kern-.75em\lower1ex\hbox{$\sim$}}\ }
\def\gl{\ \raise.5ex\hbox{$>$}\kern-.8em\lower.5ex\hbox{$<$}\ }
\def\roughly#1{\raise.3ex\hbox{$#1$\kern-.75em\lower1ex\hbox{$\sim$}}}

\begin{document}

\thispagestyle{empty}
\begin{flushright}
February 2026
\end{flushright}
\vspace*{1.5cm}
\begin{center}

{\Large \bf Cloaking Cosmic Light} 
 
\vspace*{1.3cm} {\large
Nemanja Kaloper\footnote{\tt
kaloper@physics.ucdavis.edu} }\\
\vspace{.5cm}
{\em QMAP, Department of Physics and Astronomy, University of
California}\\
\vspace{.05cm}
{\em Davis, CA 95616, USA}\\

\vspace{1.5cm} ABSTRACT
\end{center}
Light crossing dark domain walls that source a top form coupled to gauge Chern--Simons terms 
mixing visible and dark $U(1)$ gauge fields generically converts into dark photons. The effect 
is entirely localized on the wall and requires no additional ingredients. The conversion rate 
is a sharp function of the photon frequency in the wall rest frame, vanishing above the ultraviolet 
cutoff of the top form sector. Partial cloaking may also induce a rotation of the polarization of transmitted 
light of order $\Delta\vartheta \sim 10^{-3}$ radians, modify the cosmic microwave background 
power spectrum, and violate Etherington's reciprocity relation at low frequencies. 
These effects can impact cosmological determinations of the Hubble rate.

\vfill \setcounter{page}{0} \setcounter{footnote}{0}

\vspace{1cm}

\newpage

We show that standard electromagnetism can couple to dark domain walls 
through a Chern--Simons interaction that mixes it with a dark $4$-form (a.k.a top form) and a 
second $U(1)$ gauge field (a paraphoton). As a result, bubbles of evanescent 
dark energy \cite{Kaloper:2025goq}, which mediate the discrete discharge of 
the top form flux, can induce both conversion of light into dark photons and 
detectable polarization rotation, accompanied by $O(1)$ variations of the dark 
energy density. A universe containing such dark walls would not be perfectly 
transparent: a fraction of photons with frequencies below the cutoff of the
top form sector convert into paraphotons and are lost. We analyze 
the implications for localized cosmological sources and the cosmic microwave 
background (CMB), and outline possible 
consequences for determining the cosmic expansion rate.

We consider the following gauge- and Lorentz-invariant low-energy theory 
in the decoupling limit of gravity, $\mpl \to \infty$, in order to focus on gauge field
dynamics
\cite{Kaloper:2025goq,Kaloper:2025wgn,Kaloper:2025upu,Kaloper:2026slg}:
\ba
S &\ni& \int d^4 x \Bigl\{\sum_n  \Bigl(-\frac{1}{4} 
\bigl(F^{(n)}_{\mu\nu}\bigr)^2 - A^{(n)}_\mu J^{(n)~\mu} \Bigr)
- \frac{\cal H}{4! {\cal M}^2} 
\sum_{n,m} {\cal Z}_{nm} \epsilon^{\mu\nu\lambda\sigma} 
F^{(n)}_{\mu\nu} F^{(m)}_{\lambda\sigma} \nonumber \\
&&~~~~~~~~~ - \frac{1}{2} {\cal H}^2 
+ \frac{1}{6} \epsilon^{\mu\nu\lambda\sigma}  
\partial_\mu {\cal H}\, {\cal B}_{\nu\lambda\sigma} 
\Bigr\} \nonumber \\
&-&{\cal T} \int d^3 \xi \sqrt{ \bigl| \det \left(\eta_{\mu\nu} 
\frac{\partial x^\mu}{\partial \xi^a} \frac{\partial x^\nu}{\partial \xi^b} \right) \bigr| } 
- \frac{\cal Q}{6} \int d^3 \xi \, {\cal B}_{\mu\nu\lambda} 
\frac{\partial x^\mu}{\partial \xi^a} \frac{\partial x^\nu}{\partial \xi^b} 
\frac{\partial x^\lambda}{\partial \xi^c} \epsilon^{abc} \, .
\label{cantra}
\ea
Here $A^{(n)}_\mu$, $F^{(n)}_{\mu\nu}=\partial_\mu 
A^{(n)}_\nu-\partial_\nu A^{(n)}_\mu$, and $J^{{(n)}~\mu}$ 
denote the visible $(n=1)$ and dark $(n=2)$ $U(1)$ vector potentials, 
field strengths, and conserved currents, respectively. 
The pseudoscalar ${\cal H}$ is the magnetic dual of the electric top form $U(1)$ field strength 
${\cal G}_{\mu\nu\lambda\sigma}
=4\partial_{[\mu}{\cal B}_{\nu\lambda\sigma]}$, where 
${\cal B}_{\nu\lambda\sigma}$ is its electric potential and brackets denote antisymmetrization. 
The top form is sourced by membranes of tension ${\cal T}\ge 0$ and charge ${\cal Q}$. 
The mass scale ${\cal M}$ sets the strong-coupling scale of the UV completion of the top form sector.

As discussed in 
\cite{Kaloper:2025goq,Kaloper:2025wgn,Kaloper:2025upu,Kaloper:2026slg}, 
following \cite{Luscher:1978rn,Gabadadze:1997kj,Gabadadze:2002ff}, 
we take this UV completion to be a confining non-Abelian gauge theory that 
breaks chiral symmetry at a scale $\sim {\cal M}$, 
generating emergent top form dynamics \cite{Luscher:1978rn}. Thus 
${\cal M}$ is the cutoff of the low
energy theory (\ref{cantra}). 
Because the confining theory violates CP, the vacuum degeneracy is lifted 
by the discretely varying contribution ${\cal H}^2/2$, which is absent in the UV theory,
because at energies above ${\cal M}$, the terms $\propto {\cal H}$ are absent. 

The higher-rank sector in Eq.~(\ref{cantra}) originates in a dark sector, as proposed in 
\cite{Kaloper:2025goq,Kaloper:2026slg}. 
The visible and dark $U(1)$ vector fields 
could mix with each other and 
with the top form through anomaly terms involving fractionally charged dark matter,
with both their $\theta$ parameters linked to the same 
dark sector top form 
\cite{Dvali:2005an,Dvali:2005zk,Kaloper:2008qs,Kaloper:2008fb,Kaloper:2023kua}. 
The interaction is parameterized by a dimensionless matrix 
${\cal Z}_{nm}$, which generalizes the coupling $\zeta$ used in 
\cite{Kaloper:2025goq,Kaloper:2025wgn,Kaloper:2025upu,Kaloper:2026slg}. 
Since $\epsilon^{\mu\nu\lambda\sigma} F^{(n)}_{\mu\nu} F^{(m)}_{\lambda\sigma} 
= \epsilon^{\mu\nu\lambda\sigma} F^{(m)}_{\mu\nu} F^{(n)}_{\lambda\sigma}$ 
and the operator is real, ${\cal Z}_{nm}$ is real and symmetric.

We could also introduce bulk $U(1)$ mixing through kinetic bilinears, 
gauge boson masses, or fractional dark matter charges, as originally considered in 
\cite{Okun:1982xi,Georgi:1983sy,Holdom:1985ag,Nordberg:1998wn}. 
Such effects have been extensively studied and constrained by data \cite{Fabbrichesi:2020wbt}. 
For clarity we neglect them here and treat them as subleading corrections. 
Combining them with the present mechanism would be 
of interest in a more comprehensive analysis.

We now analyze the theory (\ref{cantra}). Away from the walls that source 
${\cal H}$, the bulk dynamics reduces to conventional Maxwell electrodynamics 
in both the visible and dark sectors. Varying the action with respect to $A^{(n)}_\mu$ yields
\be
\partial_\mu F^{{(n)}~\mu\nu}
= J^{{(n)}~\nu} - \frac{1}{6 {\cal M}^2}
\epsilon^{\mu\nu\lambda\sigma} \sum_m {\cal Z}_{nm} 
\partial^{~}_\mu \Bigl({\cal H} F^{(m)}_{\lambda\sigma}\Bigr) \, .
\label{mmon}
\ee
Varying (\ref{cantra}) with respect to ${\cal B}_{\nu\lambda\sigma}$ gives
\be
n^\mu \partial_\mu {\cal H} = {\cal Q} \delta \bigl(r-r(t)\bigr) \, ,
\label{Heq}
\ee
where, as in \cite{Kaloper:2026slg}, $r(t)$ denotes the wall trajectory, $n^\mu$ 
the outward-pointing normal, and $r$ the coordinate along it. 
Away from the wall, ${\cal H}$ is constant and therefore
$\frac{{\cal H}}{6 {\cal M}^2}
\epsilon^{\mu\nu\lambda\sigma}\partial^{~}_\mu F^{(n)}_{\lambda\sigma} = 0$
in the absence of visible and dark magnetic monopoles, which we assume. 
This follows from the Bianchi identities
$\partial^{~}_{[\mu}F^{(n)}_{\lambda\sigma]}=0$ in both sectors. Finally, 
varying Eq.~(\ref{cantra}) with respect to ${\cal H}$ and Hodge-dualizing yields
\be
{\cal H}_{\mu\nu\lambda\sigma}
= {\cal H}\,\epsilon_{\mu\nu\lambda\sigma}
- \frac{1}{{\cal M}^2}
\sum_{n,m} {\cal Z}_{nm} F^{(n)}_{[\mu\nu}F^{(m)}_{\lambda\sigma]} \, ,
\label{topform}
\ee
where ${\cal H}_{\mu\nu\lambda\sigma}=4\partial_{[\mu}{\cal B}_{\nu\lambda\sigma]}$ 
is the spectator electric top form, whose flux is fixed by the CP-violating membrane sources. 
Thus, off the wall the theory consists of two decoupled $U(1)$ gauge theory sectors with 
charged matter and a derivative of the mixed Chern--Simons term feeding the spectator top form 
${\cal H}_{\mu\nu\lambda\sigma}$.

The presence of walls qualitatively changes the situation. The jump of ${\cal H}$ across a 
wall by one unit of charge obstructs a global extension of the gauge field strengths 
$F^{(n)}_{\mu\nu}$ across it, as noted in \cite{Kaloper:2026slg}. The Chern--Simons 
mixing encoded by ${\cal Z}_{nm}$ modifies this structure, and complicates the 
matters slightly. However, this matrix can be diagonalized straightforwardly. 
Since ${\cal Z}_{nm}$ is real and symmetric, there exists a 
two-dimensional rotation matrix ${\cal R}(\theta)$ such that
${\cal R}^{\tt T}(\theta) {\cal Z} {\cal R}(\theta) = {\cal Z}_d$,
where ${\cal Z}_d$ is diagonal with eigenvalues $\zeta_n$, and ${\tt T}$ 
denotes transposition. Because the gauge kinetic terms are canonical, 
the kinetic matrix is unity in the $O(2)$ adjoint notation, and this rotation leaves it invariant.

We therefore change from the ``interaction'' basis $F^{(n)}_{\mu\nu}$ in (\ref{cantra}) 
to the diagonal ``propagation'' basis $f^{(n)}_{\mu\nu}$ 
that crosses the wall. Using column vectors,
\be
\begin{pmatrix}
F^{(1)}_{\mu\nu} \\
F^{(2)}_{\mu\nu}
\end{pmatrix}
=
{\cal R}(\theta)
\begin{pmatrix}
f^{(1)}_{\mu\nu} \\
f^{(2)}_{\mu\nu}
\end{pmatrix} \, .
\label{EBmatrixa}
\ee
In terms of these fields, (\ref{cantra}) becomes
\ba
S &\ni& \int d^4 x \Bigl\{\sum_n  \Bigl(-\frac{1}{4} 
\bigl(f^{(n)}_{\mu\nu}\bigr)^2 
- a^{(n)}_\mu \bigl({\cal R}^{\tt T}(\theta) J \bigr)^{{(n)}~\mu} 
- \frac{\zeta_{n} {\cal H}}{4! {\cal M}^2} 
\epsilon^{\mu\nu\lambda\sigma} f^{(n)}_{\mu\nu} f^{(n)}_{\lambda\sigma} 
\Bigr) \nonumber \\
&&~~~~~~~~~ - \frac{1}{2} {\cal H}^2 
+ \frac{1}{6} \epsilon^{\mu\nu\lambda\sigma}  
\partial_\mu {\cal H} \, {\cal B}_{\nu\lambda\sigma} 
\Bigr\} \nonumber \\
&-&{\cal T} \int d^3 \xi \sqrt{ \bigl| \det \left(\eta_{\mu\nu} 
\frac{\partial x^\mu}{\partial \xi^a} \frac{\partial x^\nu}{\partial \xi^b} \right) \bigr| } 
- \frac{\cal Q}{6} \int d^3 \xi \, {\cal B}_{\mu\nu\lambda} 
\frac{\partial x^\mu}{\partial \xi^a} \frac{\partial x^\nu}{\partial \xi^b} 
\frac{\partial x^\lambda}{\partial \xi^c} \epsilon^{abc} \, ,
\label{newcantra}
\ea
which describes two independent copies of the wall-crossing $U(1)$ sector 
studied in \cite{Kaloper:2026slg}. They are mutually decoupled when the 
charge currents $J^{(n)}_\mu$ are switched off.

Let us consider the fate of light crossing the wall. 
Since $f^{(n)}_{\mu\nu}$ are linear combinations of $F^{(n)}_{\mu\nu}$, 
both propagation eigenstates are excited even if the incident wave is a 
single interaction eigenstate. Thus an ordinary electromagnetic wave has 
two propagation channels across the wall, appearing as an admixture of 
$f^{(n)}_{\mu\nu}$ even when paraphotons are initially absent.

As in \cite{Kaloper:2026slg}, we treat the walls as ultra-thin and macroscopically large, 
approximating them as infinite planes. The wall-induced conversion in the propagation 
basis $f^{(n)}_{\mu\nu}$ is controlled by the small dimensionless parameters
$\frac{\zeta_{n} \Delta {\cal H}}{{\cal M}^2} = \frac{\zeta_{n} {\cal Q}}{{\cal M}^2}$,
which we treat perturbatively in the ``sudden" approximation. In this respect, 
the wall behaves analogously to an interface between dielectrics \cite{Jackson:1998nia}.

Since the walls are infinite, we go to their rest frame at $z=0$ and, using the boost invariance 
of the wall in the $z$--$t$ plane \cite{Huang:1985tt,Harari:1992ea}, move to the frame where 
incident waves propagate along the normal. Away from the wall, the wave dispersion relation 
is $\omega^2 = \vec k^2$ on both sides, so waves strike the wall at right angles, 
propagating in vacuum before and after.

To determine wall-induced conversion of vector gauge fields, we define 
$e^{{(n)}}_i=f^{(n)}_{0i}$ and $b^{{(n)}\,i} = \frac12 \epsilon^{ijk} f^{(n)}_{jk}$. 
It is convenient to use the redefined field strengths 
as in \cite{Kaloper:2026slg}, which, with $\sigma_n=\frac{\zeta_n {\cal H}}{3{\cal M}^2}$, are
\be
\vec {\tilde e}^{\,(n)} = \vec e^{\,(n)} - \sigma_n \vec b^{\,(n)} \, , 
\qquad 
\vec {\tilde b}^{\,(n)} = \vec b^{\,(n)} + \sigma_n \vec e^{\,(n)} \, .
\label{ebfieldsdefs}
\ee
The Maxwell equations for $J^{(n)}_\mu = 0$ read
\ba
&&\vec \nabla \cdot \vec {\tilde e}^{\,(n)}  = 0  \, , \qquad \qquad \qquad ~ \,
\vec \nabla \times \vec {\tilde b}^{\,(n)} - \partial_t \vec {\tilde e}^{\,(n)} = 0 \, ,
\label{maxeqs1}\\
&&\vec \nabla \cdot \vec {\tilde b}^{\,(n)} 
= \vec \nabla \cdot \bigl( \sigma_n \vec e^{\,(n)}  \bigr) \, , \qquad
\vec \nabla \times \vec {\tilde e}^{\,(n)} + \partial_t \vec {\tilde b}^{\,(n)} 
+ \vec \nabla \times \bigl( \sigma_n \vec b^{\,(n)} \bigr) 
- \partial_t  \bigl(\sigma_n \vec e^{\,(n)} \bigr) = 0 \, .
\label{maxeqs2}
\ea
The first two equations (\ref{maxeqs1}) correspond to $\partial_\mu \tilde f^{{(n)}~\mu\nu} = 0$, 
while (\ref{maxeqs2}) are the Bianchi identities $\partial^{~}_{[\mu} f^{(n)}_{\nu\lambda]} = 0$, 
after adding and subtracting $\vec \nabla \cdot (\sigma_n \vec e^{\,(n)})$, 
$\vec \nabla \times (\sigma_n \vec b^{\,(n)})$, and $\partial_t (\sigma_n \vec e^{\,(n)})$, 
and then using (\ref{ebfieldsdefs}) to simplify the terms under derivatives.

For $\sigma_n \ll 1$, we treat wall interactions as a perturbation of the vacuum equations. 
In vacuum, $\vec \nabla \cdot \vec e^{\,(n)} = 0$ and 
$\vec \nabla \times \vec b^{\,(n)}= \partial_t \vec e^{\,(n)}$ to leading order. 
With $\partial_t \sigma_n = 0$ and $\vec \nabla \sigma_n \parallel \vec n \parallel \vec k$, 
since the propagating waves are right-handed triads $(\vec k, \vec e^{\,(n)}, \vec b^{\,(n)})$, the 
Eqs.~(\ref{maxeqs1})--(\ref{maxeqs2}) reduce to
\be
\vec \nabla \cdot \vec {\tilde e}^{\,(n)}  = 0  \, , \quad  
\vec \nabla \times \vec {\tilde b}^{\,(n)} = \partial_t \vec {\tilde e}^{\,(n)} \, , \quad  
\vec \nabla \cdot \vec {\tilde b}^{\,(n)} = 0 \, , \quad
\vec \nabla \times \vec {\tilde e}^{\,(n)} + \partial_t \vec {\tilde b}^{\,(n)}  
= (\vec \nabla \sigma_n) \times \vec b^{\,(n)} \, .
\label{maxeqs20}
\ee        
Integrating (\ref{maxeqs20}) over Gaussian pillboxes across the wall \cite{Kaloper:2026slg}, 
the wall crossing to linear order in $\sigma_n$ imposes continuity of the redefined fields
\be
\vec n \cdot \bigl(\vec {\tilde e}^{\,(n)}_+ - \vec {\tilde e}^{\,(n)}_- \bigr) = 0 \, , ~~~
\vec n \times \bigl(\vec {\tilde b}^{\,(n)}_+ -  \vec {\tilde b}^{\,(n)}_- \bigr) = 0 \, , ~~~
\vec n \cdot \bigl(\vec {\tilde b}^{\,(n)}_+ - \vec {\tilde b}^{\,(n)}_- \bigr) = 0 \, , ~~~
\vec n \times \bigl(\vec {\tilde e}^{\,(n)}_+ -  \vec {\tilde e}^{\,(n)}_- \bigr) = 0  \, .
\label{maxeqsbcs}
\ee
Thus, wall crossing induces an {\it electromagnetic duality} transformation 
of the monopole-free $U(1)$ propagation 
eigenfields \cite{Jackson:1998nia,Shapere:1991ta}:
\be
\begin{pmatrix}
\vec e^{\,(n)}_- \\
\vec b^{\,(n)}_-
\end{pmatrix}
=
\begin{pmatrix}
1 & -\Delta \sigma_n \\
\Delta \sigma_n & 1
\end{pmatrix}
\begin{pmatrix}
\vec e^{\,(n)}_+ \\
\vec b^{\,(n)}_+
\end{pmatrix} \, ,
\label{EBmatrixo}
\ee
where
\be
\Delta \sigma_n = \frac{\zeta_n}{3{\cal M}^2} \Delta {\cal H} 
= \frac{\zeta_n}{3{\cal M}^2} {\cal Q} \, ,
\label{jumph}
\ee
is the jump across the wall due to a single unit of ${\cal H}$ flux discharge.  

In the final step we rewrite this result in terms of the gauge 
fields in the ``interaction" basis, by
combining (\ref{EBmatrixa}) and 
(\ref{EBmatrixo}). Using the 4-component column vector 
$(\vec E^{(1)}, \vec B^{(1)}, \vec E^{(2)}, \vec B^{(2)})^{\tt T}$, we obtain
\be
\begin{pmatrix}
\vec E^{(1)}_- \\
\vec B^{(1)}_- \\
\vec E^{(2)}_-\\
\vec B^{(2)}_-
\end{pmatrix}
=
{\cal P} 
\begin{pmatrix}
{\cal R}(\theta) & 0\\
0 & {\cal R}(\theta)
\end{pmatrix}
{\cal P}
\begin{pmatrix}
\Sigma_1&0 \\
0&\Sigma_2
\end{pmatrix}
{\cal P}
\begin{pmatrix}
{\cal R}^{\tt T}(\theta) & 0\\
0 & {\cal R}^{\tt T}(\theta)
\end{pmatrix}
{\cal P}
\begin{pmatrix}
\vec E^{(1)}_+ \\
\vec B^{(1)}_+ \\
\vec E^{(2)}_+\\
\vec B^{(2)}_+
\end{pmatrix} \, ,
\label{EBmatrixfin}
\ee
where ${\cal P}$ is an idempotent $4\times 4$ permutation matrix, 
${\cal R}$ the $2\times 2$ rotation (\ref{EBmatrixa}), and 
$\Sigma_n$ the wall-borne duality matrix:
\be
{\cal P} =
\begin{pmatrix}
1 & 0 & 0 & 0\\
0 & 0 & 1 & 0\\
0 & 1 & 0 & 0\\
0 & 0 & 0 & 1
\end{pmatrix} \, , 
\qquad
{\cal R}(\theta) = 
\begin{pmatrix}
\cos \theta & -\sin \theta \\
\sin \theta & \cos \theta
\end{pmatrix} \, , 
\qquad
\Sigma_n = 
\begin{pmatrix}
1 & -\Delta \sigma_n \\
\Delta \sigma_n & 1
\end{pmatrix} \, .
\label{matrices}
\ee

We now consider the implications when the incident wave is purely the 
standard electromagnetic wave from a distant source, crossing the wall 
en route to the observer. In this case, the initial dark components vanish, 
$\vec E^{(2)}_+ = \vec B^{(2)}_+ = 0$. Substituting into 
Eq.~(\ref{EBmatrixfin}) and performing the matrix multiplications, we find
\ba
\vec E^{(1)}_- &=& \vec E^{(1)}_+ - \frac{{\cal Q} 
(\zeta_2 \, \sin^2\theta + \zeta_1 \, \cos^2 \theta) }{3{\cal M}^2} \vec B^{(1)}_+ \, , \quad 
\vec B^{(1)}_- = \vec B^{(1)}_+ + \frac{{\cal Q}
(\zeta_2 \, \sin^2\theta + \zeta_1 \, \cos^2 \theta)}{3{\cal M}^2} \vec E^{(1)}_+ \, , \nonumber \\
\vec E^{(2)}_- &=& \frac{{\cal Q} 
\sin\theta  \cos\theta (\zeta_2 - \zeta_1)}{3{\cal M}^2} \vec B^{(1)}_+ \, , \quad
\qquad \qquad \,
\vec B^{(2)}_- = - \frac{{\cal Q} 
\sin\theta  \cos\theta (\zeta_2 - \zeta_1)}{3{\cal M}^2} \vec E^{(1)}_+ \, .
\label{EB-+}
\ea
The first line shows that the transmitted 
electromagnetic wave experiences polarization rotation by
\be
\Delta \vartheta \simeq 
\frac{{\cal Q}(\zeta_2 \, \sin^2\theta + \zeta_1 \, \cos^2 \theta)}{3{\cal M}^2} \ll 1 \, ,
\label{rotangle}
\ee
generalizing \cite{Kaloper:2026slg} by including contributions from the dark 
$U(1)$ channel. Applying this to the CMB, polarization rotation analyses imply 
that $\Delta \vartheta$ is at most a degree of angle, or so \cite{Komatsu:2022nvu}.

The second line of Eq.~(\ref{EB-+}) shows that a fraction of the visible 
wave converts into dark $U(1)$ during crossing. The fractional loss is 
estimated by comparing the transmitted dark intensity with the initial 
visible intensity. Since the waves are right-handed triads, their intensities 
are reliably approximated by our equations to quadratic order in $\sigma_n$. 
To this order, the transmitted dark power fraction is
\be
\gamma \simeq \Bigl(\frac{\cal Q}{3{\cal M}^2}\Bigr)^2 
\sin^2 \theta \cos^2 \theta (\zeta_2 - \zeta_1)^2 \, .
\label{wavedec}
\ee
We ignore the small apparent amplification of the visible wave, suggested by formulas 
(\ref{EB-+}), that indicate a slight increase of wave intensity upon crossing.
This behavior is a known pathology in wave scattering, 
where the first-order perturbation fails by suggesting that the
probability that an initial state scatters into itself exceeds unity. Because the wave 
fields in (\ref{EB-+}) form right-handed triads, this excess
however is ${\cal O}(\sigma^4)$ and can be safely ignored. The issue 
is fully resolved in Brillouin-Wigner approach to perturbation theory, 
by imposing unitarity, calculating off-diagonal 
transitions, and then subtracting those rates from unity.

From Eq.~(\ref{wavedec}), if $\zeta_1$ and $\zeta_2$ have the same 
sign, visible-to-dark transitions are constrained by polarization bounds, 
$\gamma \sim \Delta \vartheta^2 \lesssim 10^{-6}$. An interesting possibility 
arises if $\zeta_1$ and $\zeta_2$ have opposite signs and 
$\tan\theta \simeq \pm \sqrt{|\zeta_1/\zeta_2|}$, suppressing 
$\Delta \vartheta$ while enhancing the conversion:
\be
\gamma \simeq \Bigl(\frac{\cal Q}{3{\cal M}^2}\Bigr)^2 \zeta_2^2 \tan^2 \theta 
\simeq \Bigl(\frac{\cal Q}{3{\cal M}^2}\Bigr)^2 |\zeta_1 \zeta_2| \, ,
\label{gammaspec}
\ee
which can be appreciably larger than $\Delta \vartheta^2$. In this regime, 
sources of light behind the wall could appear significantly dimmed, potentially 
with $\Delta \vartheta \sim 10^{-3}$ and $\gamma$ of order a few percent.

To estimate the visible power loss in wall crossing, Eq.~(\ref{wavedec}), 
we ignored contributions from incident dark waves. This is reasonable for 
sources composed of the Standard Model particles, which are not prolific 
emitters of dark light. One might wonder about what happens with the CMB, 
since in other models \cite{Okun:1982xi,Georgi:1983sy,Holdom:1985ag,Nordberg:1998wn} 
visible-dark mixing may induce distortions of the CMB spectrum, which can be 
constrained by observations \cite{Fabbrichesi:2020wbt}.

As an illustrative example, consider photon-axion mixing \cite{Csaki:2001yk} affecting 
Type Ia supernovae used to probe dark energy, reviewed in \cite{Mirizzi:2006zy}. 
There, ultralight axion-like fields mix with light 
in the presence of large scale, randomly oriented 
cosmic magnetic fields. Because of the randomness, the mixing is not coherent but
stochastic, leading to flavor equilibration. The effect is cumulative, 
frequency-dependent, and requires very specific environmental conditions --
nano-Gauss magnetic fields, very dilute extragalactic plasma, 
and absence of axion background -- to be significant.
Even under optimal conditions, the mixing 
may induce quasar color shifts or CMB distortions \cite{Mirizzi:2006zy}. The limits on these distortions
indicate that while the photon-axion mixing can affect the 
supernovae observation results, it cannot be the dominant source of 
supernova dimming. Other mixing mechanism in the bulk experience similar limitations. 

In contrast, the wall-localized Chern-Simons mixing of visible and dark 
electromagnetism is much stealthier. A wall with sub-MeV tension  
does not produce significant gravitational anisotropies \cite{Zeldovich:1974uw}. 
The wall-wrapped bubbles will not be produced in very large electromagnetic 
fields when the UV cutoff is low \cite{Kaloper:2026slg}. 
The mixing is completely localized to the wall; it is neither a continuous flavor 
oscillation nor a frequency-dependent cumulative effect. 
A single wall produces a fixed power loss incurred only as light crosses it, 
in the frequency range bounded from above by the cutoff  ${\cal M}$ of the top form sector.

For frequencies above the cutoff ${\cal M}$ the wall becomes transparent. 
The visible-to-dark conversions at high frequencies do not occur 
since those waves propagate through the wall like gamma-rays through 
ordinary matter; in fact, even more easily. We can infer this in 
discretely evanescent dark energy \cite{Kaloper:2025goq}, 
where the UV completion is an asymptotically free 
gauge theory with degenerate vacua and no Chern-Simons 
walls above the cutoff. Below the cutoff, flavor-mixing 
terms $\sim \vec E^{(1)} \cdot \vec B^{(2)}$ 
and the like (as needed to maintain gauge symmetry and Lorentz 
symmetry) intertwine the incoming states once they hit 
the wall. Off the wall, these 
terms are total derivatives and do not affect local fields.

On a wall with Chern-Simons terms, the jump in ${\cal H}$ 
enforces a single, localized mixing event. In fact, 
even polarization rotation can be viewed as a 
flavor mixing of photon helicities. If polarization rotation is 
sufficiently small to satisfy the CMB bounds \cite{Komatsu:2022nvu},
\be
\Delta \vartheta \sim 10^{-3}\,{\rm radians} \, ,
\label{variations}
\ee
the visible-to-dark mixing could yield observable effects. As long as such 
walls are rare and have not percolated or collided extensively at small redshifts, 
low-tension walls could still be around, 
having evaded detection to date \cite{Koren:2025ymq}.

In \cite{Kaloper:2025goq} we considered a cutoff ${\cal M} \sim {\rm milli\!-\!eV}$ for 
simplicity. In that framework, the cloaking of the visible light would only affect
frequencies $\omega \la {\rm milli\!-\!eV}$; this means a fraction of 
the CMB would be converted to dark light, while optical range sources would
remain unaffected. Minor modifications of the model in \cite{Kaloper:2025goq} 
allow ${\cal M} \sim {\rm eV}$, coinciding with the 
scale of early dark energy \cite{Poulin:2018cxd}. 
In such a scenario, the wall can mix visible and dark 
waves with optical frequencies $\sim {\rm eV}$ and below, 
while remaining transparent to ultraviolet waves and beyond.
In such a framework one could also simultaneously incorporate both early and late 
dark energy if the late vacuum energy was not fully cancelled, e.g. for 
$\sqrt{\cal Q}/{\cal M} < 1$. In this work, we will be agnostic to 
which option is preferred.

As we noted above, we have neglected dark sector 
contributions to the incident waves from sources behind the 
wall, which is reasonable for sources made 
up of Standard Model particles. On the other hand the early universe dynamics 
generically produces a relic dark radiation background due to inflation, reheating, and dark matter 
interactions. Cosmological constraints, particularly from BBN and late-time observations, 
limit these contributions to be subleading. In \cite{Kaloper:2025goq} we noted that early dark 
radiation in our case is generically colder than the early 
CMB, with $T^{\tt dark}/T^{\tt visible} \la 0.001 - 0.01$. 
Consequently, the dark photon number is extremely suppressed in the 
CMB frequency range ($\omega \sim 0.1-1~{\rm meV}$),
\be
\frac{N(\omega)_{\tt dark}}{N(\omega)_{\tt visible}} \Big|_{\rm max} 
\simeq e^{-\omega/T^{\tt dark}} < 10^{-5} \, ,
\label{dvsupr}
\ee
so the only relevant mixing process is on-the-wall conversion of CMB photons into 
dark photons. Being frequency-independent below the cutoff ${\cal M}$, the 
wall creates a dark ``snapshot" of the CMB, depleting the visible CMB by 
$\rho = 1-\gamma$, with $\gamma$ from Eq.~(\ref{wavedec}). In our approximation, where
we ignored cosmic expansion, for a large wall or bubble 
the depletion does not induce significant corrections to anisotropies.
The main effect is a rescaling of the CMB black body spectrum at wall crossing,
$N(\omega) d\omega = \frac{\omega^3 d\omega}{4\pi^3} (e^{\omega/T}-1)^{-1}$,
by a factor of $\rho$. 
Due to the form invariance of the CMB photon 
number $N(\omega)_{\tt visible} \, d\omega$ under rescaling
of frequency, the factor $\rho$ can be absorbed into the measured 
temperature inside the bubble, 
\be
T^{\tt visible}_- = \rho^{1/4} \, T^{\tt visible}_+ \, .
\label{tren}
\ee
This shifts the last scattering surface of the 
CMB as seen by a terrestrial observer. 
The wall-induced depletion of the CMB power thus allows 
the CMB to reach the observed temperature more efficiently, 
needing less time and space to ``cool" down to the observed temperature.
If we combine this with cosmic expansion, the CMB temperature would decrease 
in less time, that could increase $H_0$, due to the 
scaling invariance described above. The scaling remains an approximate 
degeneracy between $H_0$ and $T^{\tt visible}_0$ 
in an expanding universe \cite{Ivanov:2020mfr}. 

After crossing the wall, the lost CMB photons 
convert into the dark ones, replicating 
the CMB spectrum at the crossing. 
Their power is given by $\gamma$ times the original
CMB power. We can interpret this as the 
generation of the dark thermal background on our 
side of the wall, with the temperature 
$T^{\tt dark}_- = \gamma^{1/4} \, T^{\tt visible}_+$, 
similarly to (\ref{tren}). 
Observational constraints on the $T_{\tt CMB}$ evolution \cite{Hurier:2013ona} 
are precise to $\sim 0.5\%$ at low redshifts and a few percent at $z\sim 0.65$, with 
the fitting of evolution done relative to $\Lambda CDM$. 
While these bounds are pretty tight, a mere $2\%$ variation of $T_{\tt CMB}$  
yields a power reduction of $(0.98)^4 \sim 0.92$, 
or $\sim 8\%$, induced by the wall-borne Chern-Simons mixing. 
Correspondingly the dark background temperature right after the wall crossing would be
$T^{\tt dark}_-/T^{\tt visible}_+ \simeq \gamma^{1/4}$, that could be as much as $0.5$. 
This might contribute to $\Delta N_{\tt eff}$, but only at very low redshifts, 
much later than recombination.

In the case where ${\cal M} \sim {\rm few~eV}$, 
the depletion of visible light by conversion to
dark photons by our mechanism would also affect optical range sources. 
The light-to-dark conversion could reduce 
optical flux from sources behind the wall by the same
fraction as the CMB, due to its frequency independence below the cutoff ${\cal M}$. This could  
additionally impact the inferred Hubble rate from the CMB 
versus low-redshift sources \cite{Ivanov:2020mfr,Kamionkowski:2022pkx}.

Such attenuation may also lead to violations of the Etherington reciprocity 
relation, which relates luminosity and angular diameter distances via 
$D_L(z)/D_A(z) = (1+z)^2$ \cite{Etherington:1933asu,Bassett:2003vu}. 
While most tests probe smooth, cumulative violations of this ratio, 
step-like luminosity changes at specific redshifts have been considered recently 
\cite{Teixeira:2025czm} as a possible way to address the $H_0$ tension. 
In our scenario, the violation would be step-like location-wise, at a specific redshift with  
a correlated change in vacuum energy due to the top form flux discharge. 
It would also only affect the light with frequencies below the cutoff 
${\cal M}$ of the dark top form sector, rapidly shutting off at higher frequencies. 

In summary, Chern-Simons terms mixing visible and dark $U(1)$ vector gauge fields 
on discretely evanescent dark energy walls can alter the apparent brightness 
of sources behind the wall. The jump in luminosity is correlated with a vacuum 
energy jump. The luminosity jump could be interpreted as a shift of the
last scattering surface, after redefining the CMB temperature on the far side of the wall. 
The mechanism is stealthy, potentially evading 
most current probes. Observations of the CMB polarization rotation 
\cite{Komatsu:2022nvu} and the $H_0$ tension \cite{Kamionkowski:2022pkx}
may offer the first hints of such 
dark sector physics, motivating further investigation of these exotic phenomena.

\vskip.3cm

{\bf Acknowledgments}: We thank A. Westphal and especially G. D'Amico for discussions. 
This research was supported in part by the DOE Grant DE-SC0009999.

\end{document}